\documentstyle{europhys}

\def\And{{\rm and\ }}

\def\stars{\bigskip\centerline{***}\medskip}

\newif\ifboo \boofalse

\begin{document}
\euro{}{}{}{}
\Date{}
\shorttitle{}
\title{Paramagnetic state in $d$-wave Superconductors}
\author{H. Won\inst{1}, H. Jang\inst{1}
\And K. Maki\inst{2}}
\institute{
\inst{1} Department of Physics and IRC, Hallym University, 
Chunchon 200-702, South Korea \\
\inst{2}Department of Physics and Astronomy,
 University of Southern California, Los Angeles, CA 90089-0484, USA}
\rec{}{}
\maketitle
\begin{abstract}
We study theoretically the paramagnetic state in $d$-wave 
superconductors.
We present the specific heat, the magnetization, 
superfluid density obtained 
within the weak-coupling model. At low
temperatures and for small magnetic fields
they exhibit simple power law behaviors,
which should be accessible experimentally  
in hole-doped high-$T_c$  cuprates 
and $\kappa$-(ET)$_2$ salts in  a magnetic field within 
the conducting plane.
\end{abstract}
Perhaps the most momentous event in the recent history of superconductivity
is the identification of $d$-wave
symmetry in the hole-doped high-$T_c$ cuprates
superconductors \cite{tsuei_1,maki_ann}.
We believe now most of newly discovered 
superconductors; heavy fermion supercondutors 
and organic superonductors are unconventional \cite{won_klu},
though of course these possibilities have been
widely discussed in the literature \cite{heffner,lang}.

\par
First we would like to point
out that the superconductivity in
$\kappa$-(ET)$_2$ salts like the one in 
$\kappa$-(ET)$_2$Cu(N(CN)$_2$)Br is most likely
of $d_{xy}$-wave \cite{pinteric}.
Clearly earlier $T_{1}$ measurement of $^{13}$C NMR \cite{mayaffre},
the specific data\cite{nakazawa},
and the more recent thermal conductivity data  \cite{belin}
as well as a recent measurement of the magnetic penetration
depth  in the superconducting state of $\kappa$-(ET)$_2$X
\cite{pinteric} 
support that the superconductivity is of $d_{xy}$-wave.
\par
As is well known both high-$T_c$ cuprates and $\kappa$-(ET)$_2$salts
have layered structure with weak interlayer coupling.
One of the signatures of the weak interlayer coupling and $d$-wave
superconductivity is the peak in the
out-of-plane magnetoresistance just before the superconducting transition
\cite{won_klu,maki_synth}
as seen by Ito et al. \cite{ito}
and Kartsovnik et al. \cite{kartsovnik}.
In these systems, we believe  that the orbital effect is secondary when
a magnetic field is applied
parallel to the conducting plane. In such a situation the 
Pauli paramagnetism should
be of prime importance  \cite{clogsten}.
In order to test this idea we study theoretically the paramagnetic
state in $d$-wave superconductors.
Perhaps $\kappa$-(ET)$_2$salts will 
provide an ideal system to test the theoretical prediction.
\par
In this Letter we shall neglect deliberately the question of the
nonhomogeneous superconducting state in $d$-wave superconductors
\cite{maki_czech,yang} analogous to the one
discussed in $s$-wave superconductors \cite{fulde,larkin},
but rather concentrate on the paramagnetic state with
uniform order parameter.
Therefore our main purpose is to 
extend earlier works on $s$-wave superconductors 
\cite{sarma,maki_prog} to $d$-wave superconductors.
Indeed the presence of nodes in $d$-wave order parameter
gives rise to a number of new effects, some of which have been 
already discussed in \cite{yang}.
In the following
we take the weak-coupling model for $d$-wave superconductors
\cite{won_phyb} and study the 
paramagnetic state.
First of all the gap equation is given by
\begin{equation}
1=\lambda 
\displaystyle\int_{0}^{E_{c}} d \xi 
<\frac{\cos^2 2 \phi}{E}
(\tanh \frac{E+h}{2T} 
+ \tanh \frac{E-h}{2 T})>
\end{equation}
where  $E =\sqrt{\xi^2 + \Delta^2 \cos^2 2\phi}$, 
$h=\mu_{B} B$ with
the
$\mu_{B}$ the Bohr magneton, and $\lambda$ is the dimensionless
coupling
constant.
Also $< \ldots > $means the average over $\phi$.
The above equation is transformed into
\begin{equation}
-\ln(\frac{\Delta}{\Delta_{00}}) = \displaystyle\int_0^{\infty} dx
\Phi(x)[\frac{1}{1+e^{\beta(\Delta x + h)}} +
\frac{1}{1+e^{\beta(\Delta x - h)}} ]
\end{equation}
where
\begin{eqnarray}\nonumber
\Phi(x) & = & \frac{2}{\pi}( K(x) - E(x) )\qquad \qquad {\rm for} 
                                         \quad x \leq 1 \\
        & = & \frac{2}{\pi}x( K(\frac 1x) - E(\frac 1x))\qquad\qquad {\rm for}
                 \quad x > 1
\end{eqnarray}
and $\Delta_{00}$ is the superconducting order parameter for 
$T=0$ and $h=0$, 
$\Delta_{00} = 2.14 T_{c0}$, 
$T_{c0}$ the transition temperature at
$h=0$, and $K(x)$ and $E(x)$ are the complete elliptic integrals.
\par
For $T,\,\,  h \, <<\Delta_{00}$, Eq.(2) can be expanded as 
\begin{eqnarray}
-\ln(\frac{\Delta}{\Delta_{00}}) 
& \simeq & 3\zeta(3)(\frac{T}{\Delta})^3
+ 2\ln2(\frac hT)^2 + \frac{1}{24}(\frac hT)^4 + ...
\quad {\rm for}\quad \frac hT \ll 1 \\
&\simeq & \frac{1}{3}(\frac{h}{\Delta})^3 +
 \frac{\pi^2}{6}(\frac{h}{\Delta})^3(\frac{T}{h})^2
+\ldots
\qquad \qquad\qquad{\rm for}\quad \frac hT \gg 1
\end{eqnarray}

\begin{figure}
\caption{(a) The order parameter in $d$-wave superconductor
is shown as a function of $T/T_{c0}$ for several fields.
(b) The order parameter in $d$-wave superconductor
is shown as a function of $h/\Delta_{00}$ for several temperatures.}  
\end{figure}

Eq.(2) is solved numerically and shown in Fig.1 (a) and(b).
The asymptotic behavior for $\Delta$, 
Eq.(5), has been mentioned already in \cite{yang}.
Making use of $\Delta$ obtained in Eq.(2) the free energy of the 
superconducting state
is given by
\begin{eqnarray}\nonumber
\Omega_{s} &=&
-\frac14 N_0\Delta^2( 1 + 2 \ln \frac{\Delta_{00}}{\Delta})
-\frac{2N_0}{\beta} \int_{0}^{E_{c}} \!\! d \xi 
\big< \ln (1 + e^{-\beta(E+h)}) + \log (1 + e^{-\beta(E-h)})
\big>
  \\
&=& -N_0\Delta^2 [\frac14 + \displaystyle\int_0^{\infty} dx A(x)
[\frac{1}{1+e^{\beta(\Delta x + h)}} +
\frac{1}{1+e^{\beta(\Delta x - h)}} ] ]
\end{eqnarray}
where
\begin{eqnarray}\nonumber
A(x)& = &\frac{2}{\pi} [ (2x^2 - 1)K(x) + E(x) ]
\qquad\qquad{\rm for}\quad x \leq 1\\
& = &\frac{2}{\pi} x ( K(\frac 1x) + E(\frac 1x))
\qquad\qquad\qquad\qquad{\rm for}\quad x > 1
\end{eqnarray}

\begin{figure}
\caption{(a) The free energy in $d$-wave superconducting state
is shown as a function of $T/T_{c0}$ for  several fields.
(b) The free energy in $d$-wave superconducting state 
is shown as a function of $h/\Delta_{00}$ for several temperatures.
The dotted lines represent the free energy in the metastable state.}  
\end{figure}

We show in Fig.2 (a) and (b),  $\Omega_s/(N_{0}\Delta_{00}^2/4)$
 as function of 
$h/\Delta_{00}$ for several $T/T_{c0}$  and  as function of $T/T_{c0}$
for several $h/\Delta_{00}$, respectively.
For $T, h << \Delta_{00}$,
\begin{equation}
\Omega_{s} = -N_0\Delta^2(\frac14 + \frac92\zeta(3)(\frac{T}{\Delta})^3
+3 \ln2(\frac{T}{\Delta})^3(\frac hT)^2 + . . .) \qquad\qquad{\rm for}\qquad\frac hT \ll 1
\end{equation}
\begin{equation}
= -N_0\Delta^2(\frac14 + \frac12(\frac{h}{\Delta})^3 + 
\frac{\pi^2}{2}(\frac{T}{\Delta})^3(\frac{h}{T}) +
O(e^{-h/T}))
\qquad\qquad{\rm for}\qquad\frac hT \gg 1
\end{equation} 

\begin{figure}
\caption{(a) The magnetization  
is shown as a function of $T/T_{c0}$ for several fields.
(b) The magnetization 
is shown as a function of $h/\Delta_{00}$ for  several temperatures.}
\end{figure}

Similarly the magnetization (due to spin) and the specific heat are given by
\begin{equation}
M = -2N_0\Delta\displaystyle\int_0^{\infty} dx \frac{N(x)}{N_0}
( \frac{1}{1+e^{\beta(\Delta x + h)}} -
\frac{1}{1+e^{\beta(\Delta x - h)}} )
\end{equation}
and
\begin{eqnarray}\nonumber
C_{s}&=& 2\beta^2N_0\displaystyle\int_0^{\infty} dE [
f(E+h)(1-f(E+h))\{(E+h)^2\frac{N(\frac{E}{\Delta})}{N_0} - 
\frac12\frac{E+h}{E}T\frac{\partial\Delta^2}{\partial T}
R(\frac{E}{\Delta})\}  \\
& & \quad\quad\quad\qquad+\quad ( h\,\,\, \longrightarrow\,\,\, -h) \,\,\,\,
 ]
\end{eqnarray}
where $N(x)$ and $R(x)$ are given by
\begin{eqnarray}\nonumber
N(x)/N_0 & = \displaystyle \frac{2}{\pi} x K(x)\qquad\qquad {\rm for}\quad x \leq 1 \\
& = \displaystyle \frac{2}{\pi}K(\frac 1x) \qquad\qquad {\rm for}\quad x > 1
\end{eqnarray}
and
\begin{eqnarray}\nonumber
R(x) & =  {\displaystyle\frac{2}{\pi}} x [K(x) - E(x)] 
\qquad\qquad {\rm for}\quad x \leq 1 \\
& =  {\displaystyle \frac{2}{\pi}} x^2 [K(\frac 1x) - E(\frac 1x)] 
\qquad\qquad {\rm for}\quad x > 1
\end{eqnarray}
and $N(x)$ is the quasi-particle density of states.
$M/N_{0}\Delta_{00}$ is shown in Fig. 3 (a) and (b), while
$C_{s}/(\frac{\pi^2}{3})N_{0}T_{c}$
is shown in Fig. 4(a).  
For $T, h << \Delta_{00}$ we obtain
\begin{eqnarray}\nonumber
M \simeq & N_0\Delta
[ \displaystyle 4\ln2  (\frac{T}{\Delta})^2(\frac{h}{T}) +
\frac{1}{6} (\frac{T}{\Delta})^2(\frac{h}{T})^3 + ...]
\qquad {\rm for}\quad \frac hT \ll 1 \\
\simeq & N_0\Delta[\displaystyle 
(\frac{h}{\Delta})^2 +\frac{\pi^2}{3}(\frac{T}{\Delta})^2 
+  O(e^{-h/T}) ] \qquad\qquad {\rm for}\quad \frac hT \gg 1
\end{eqnarray}
and
\begin{eqnarray}\nonumber
C_{s} & \simeq & 
18 \zeta(3)\frac{N_{0}T^2}{\Delta} \big[
 1 +  O(\frac hT)^4 \big]
\qquad\qquad\qquad\qquad\qquad {\rm for}\quad \frac hT \ll 1  \\
& \simeq & \frac{2\pi^2}{3}N_{0}\frac{Th}{\Delta}  \big[ 1 + O( e^{-h/T}) \big]
\qquad\qquad\qquad\qquad\qquad\qquad {\rm for}\quad \frac hT \gg 1
\end{eqnarray}
The jump in the specific heat is given by 
\begin{equation}
\Delta C_s  /C_{N}(T_{c}(h))  = \displaystyle 
-\frac{16}{Re[\psi^{(2)}(\frac12 
+\frac{ih}{2\pi T_c(h)} ) ]}
(1+{\rm Im}(\psi^{(1)}(\frac12+\frac{ih}{2\pi T_c(h)}))
\frac{h}{2\pi T_c(h)})^2
\end{equation}
We note that it becomes sharper as we approach
$h \sim 0.501 \Delta$ where the order of the transition 
changes to the first order
\cite{maki_prog}. It is shown in Fig. 4(b).

\begin{figure}
\caption{(a) The specific heat is shown as a function of temperature
for several fields.
(b) The jump in the specific heat is shown as
a function of fields. It diverges as the field approaches
to 0.501 $\Delta_{00}$.
}
\end{figure}

\par
The in-plane and the out of plane superfluid density are given by
\begin{eqnarray} \nonumber
\frac{\rho_{s,{\rm in}}(T,h)}
{\rho_{s,{\rm in}}(0,0)} 
&=& 1 + 2 \int_{0}^{\infty} d\xi <\sin^2\phi 
(\frac{\partial f(E+h)}{ \partial E} + \frac{\partial f(E-h)}{ \partial E})
> \\ &=& 1 - \frac 1T\displaystyle\int_0^{\infty} dE 
\frac{N(E/\Delta)}{N_0}[{\rm sech}^2\frac{E+h}{2T} +
{\rm sech}^2(\frac{E-h}{2T})]
\end{eqnarray}
and
\begin{equation} 
\frac{\rho_{s,{\rm out}}(T,h)}
{\rho_{s,{\rm out}}(0,0)}
=
\frac{\Delta}{\Delta_{00}}
\frac{\pi}{2}\displaystyle
<\cos 2\phi \tanh \frac{\Delta \cos2\phi +h}{2T}>
\end{equation}
Here we used Ambegaokar-Baratoff formula for Josephson 
current \cite{ambegaokar} to calculate the out-of-plane superfluid density.
At T = 0K, Eq(17) and Eq(18) reduces to
\begin{equation}
\frac{\rho_{s,{\rm in}}(0,h)}{\rho_{s,{\rm in}}(0,0)} = 
1 - \frac{2}{\pi}\frac{h}{\Delta}K(\frac{h}{\Delta})
\end{equation}
and
\begin{equation}
\frac{\rho_{s,{\rm out}}(0,h)}{\rho_{s,{\rm out}}(0,0)} = 
\sqrt{1 - (\frac{h}{\Delta})^2}
\end{equation}
For h,T $\ll \Delta_{00}$ , we have
\begin{eqnarray}\nonumber
\frac{\rho_{s,{\rm in}}(T,h)}{\rho_{s,{\rm in}}(0,0)} 
&\simeq& 1 - 
2\ln2(\frac{T}{\Delta}) - \frac14(\frac{T}{\Delta})(\frac hT)^2
+ ... \quad\quad
\qquad\qquad {\rm for}\quad \frac hT << 1 \\
&\simeq&1 - \frac{h}{\Delta} - 2(\frac{h}{\Delta})(\frac Th)e^{-\frac hT}
+O(e^{-2h/T})
\qquad\qquad {\rm for}\quad \frac hT >> 1
\end{eqnarray}
and
\begin{equation}
\frac{\rho_{s,{\rm out}}(T,h)}{\rho_{s,{\rm out}}(0,0)} 
\simeq
\sqrt{1 - (\frac{h}{\Delta})^2}
-  \frac{\pi^2}{6}(\frac{T}{\Delta})^2 
-\frac{\pi^2}{4}(\frac{T}{\Delta})^2(\frac{h}{\Delta})^2
+ O( (\frac{h}{\Delta})^4, (\frac{T}{\Delta})^4)
\end{equation}
In the last case the same expression applies for $0\le h/T < \infty$.
$\rho_{s,{\rm in}}(T,h)$ and $\rho_{s,{\rm out}}(T,h) $ are shown in Fig.5 
and 6, respectively.

\begin{figure}
\caption{(a) The superfluid density in the plane 
is shown as a function of $T/T_{c0}$ for  several fields.
(b) The superfluid density in the plane  
is shown as a function of $h/\Delta_{00}$ for several temperatures.}  
\end{figure}

\begin{figure}
\caption{(a)The $c$-axis superfluid density  
is shown as a function of $T/T_{c0}$ for  several fields.
(b) The $c$-axis superfluid density 
is shown as a function of $h/\Delta_{00}$ for several temperatures.}  
\end{figure}

\begin{figure}
\caption{ The quasi particle density of states in the superconducting state 
is shown as a function of field and temperature.}
\end{figure}

\par 
Finally the quasi-particle density of states in a magnetic field is 
shown in Fig. 7.
In particular the density of states exhibits a flat portion which extend
for $|E|< h$. Also the peaks at $E =\pm \Delta$ splits into double peaks at 
$E =\pm(\Delta \pm h)$.

\par
In summary we have studied the paramagnetic state in $d$-wave superconductors,
which may be realized in high-$T_c$ cuprates
and $\kappa$-(ET)$_2$salts in the presence of a magnetic field within the 
conducting plane
(the $a$-$b$ plane and the $a$-$c$ plane, respcetively) .
For simplicity we have neglected the orbital effect associated 
with magnetic field.
However in many of above systems we believe the paramagnetic 
effect is predominant.
In this situation the most properties 
in the above systems exhibit simple power laws as
we have described here, which should be
readily accessible experimentally.
Therefore  the observation of these  properties 
described  should provide  another signature  of 
$d$-wave superconductivity in these systems.

\stars
H. Won acknowledges support 
in part from the Korea Science and engineering
under grant number 96-0702-02-01-3 and
in part from the Korea Research Foundation
under grant number 1998-015-D00055.
The present work is also supported by the National Science Foundation under
grant number DMR95-31720.

\vskip-12pt

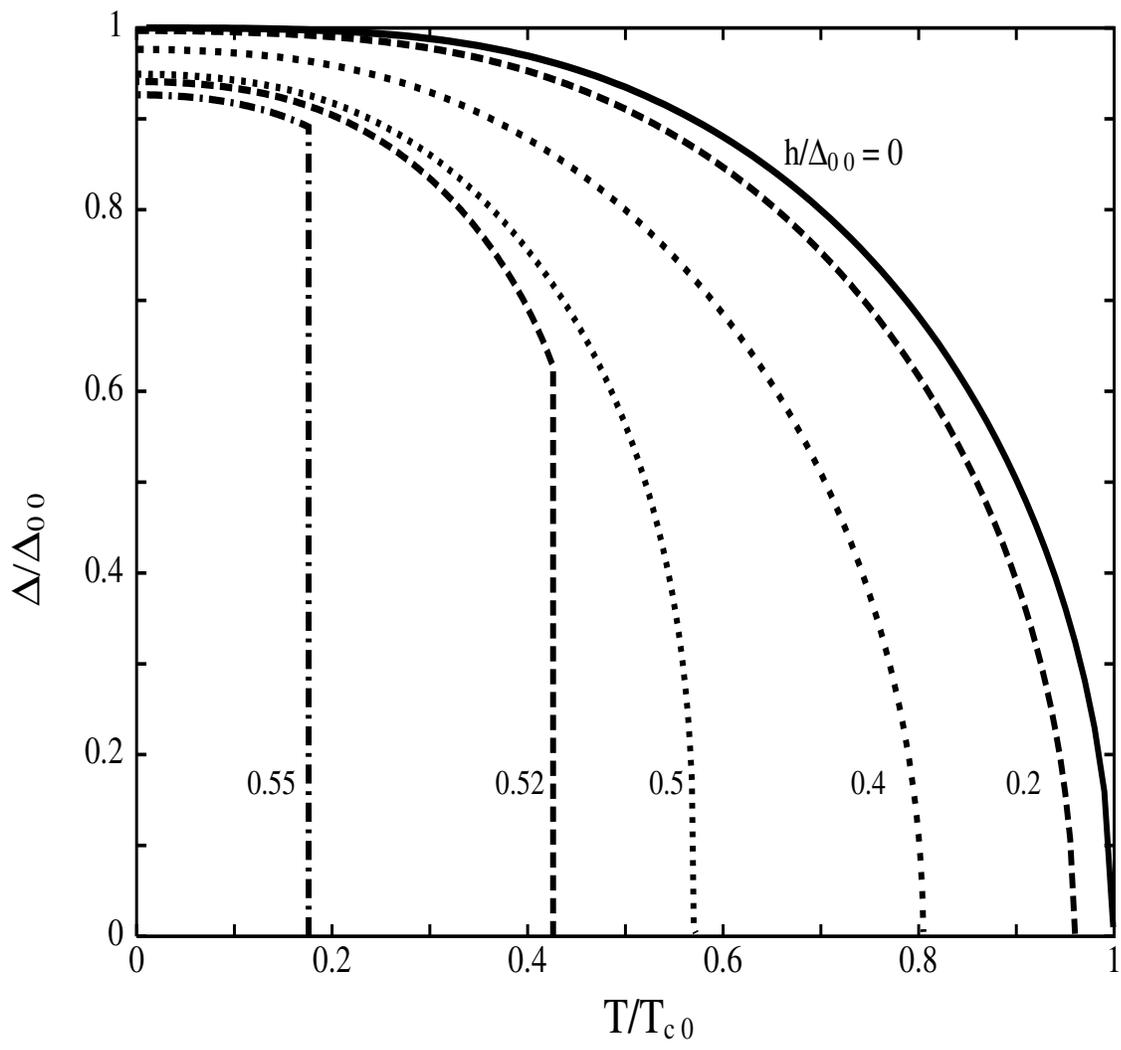
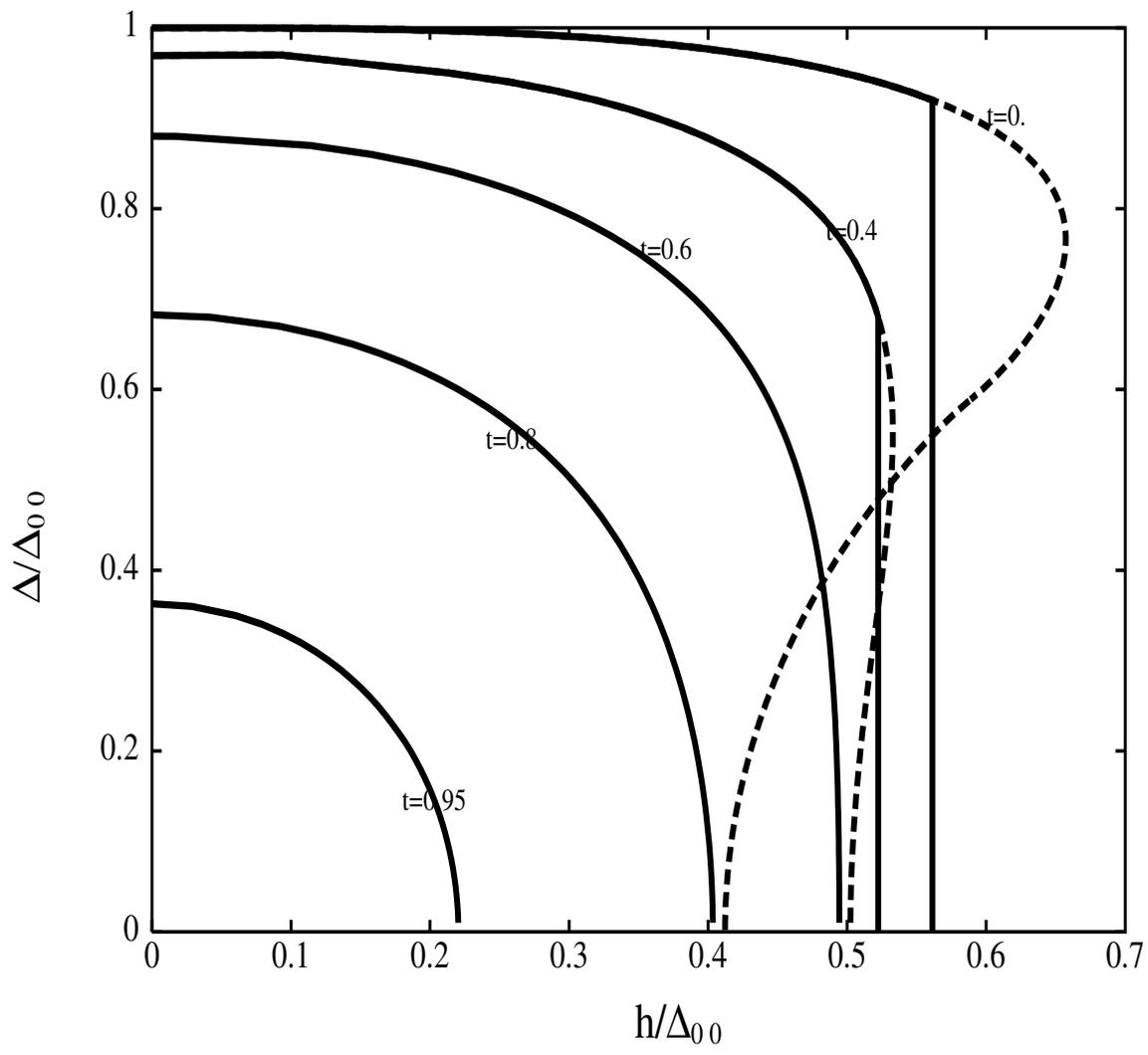
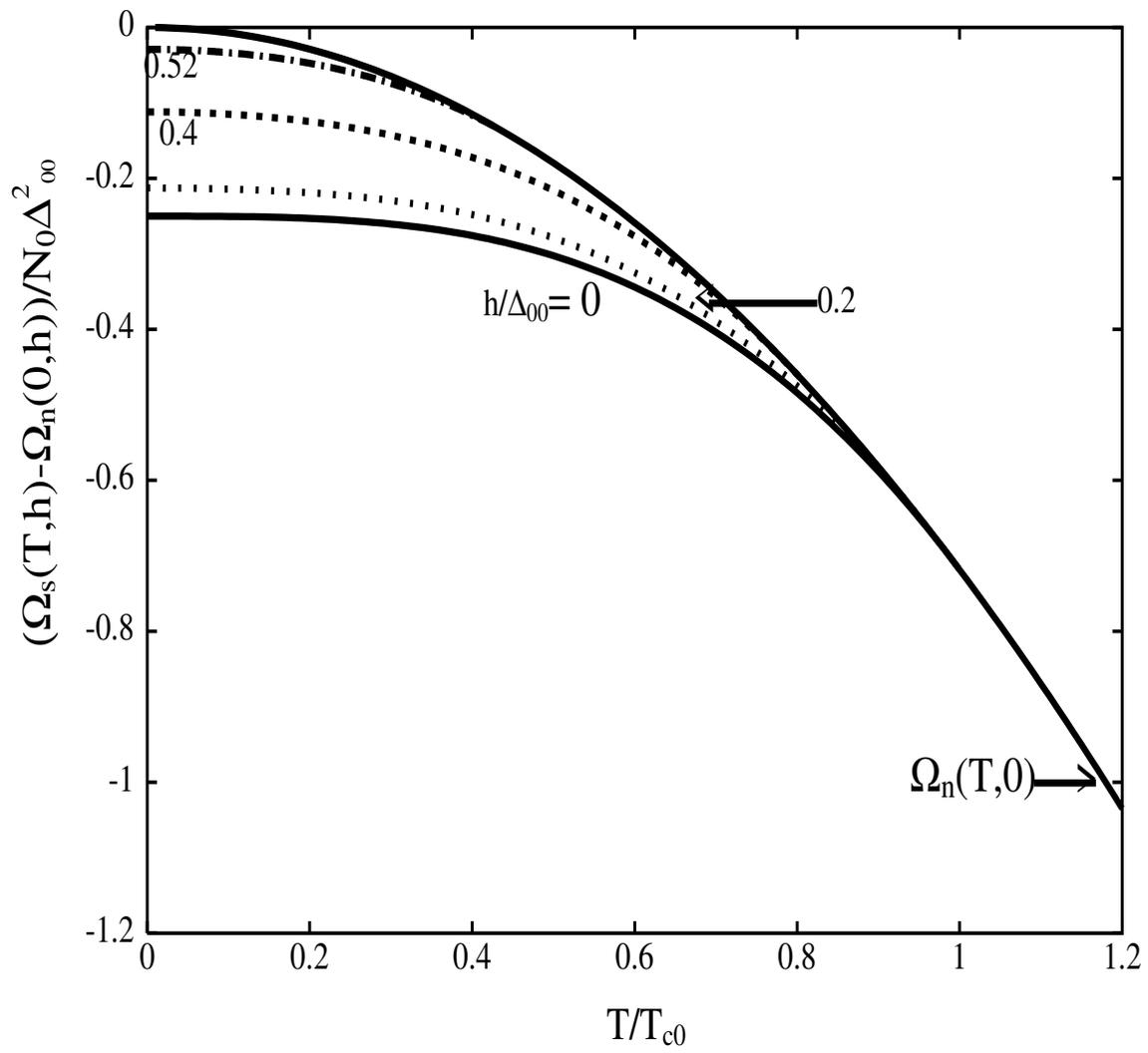
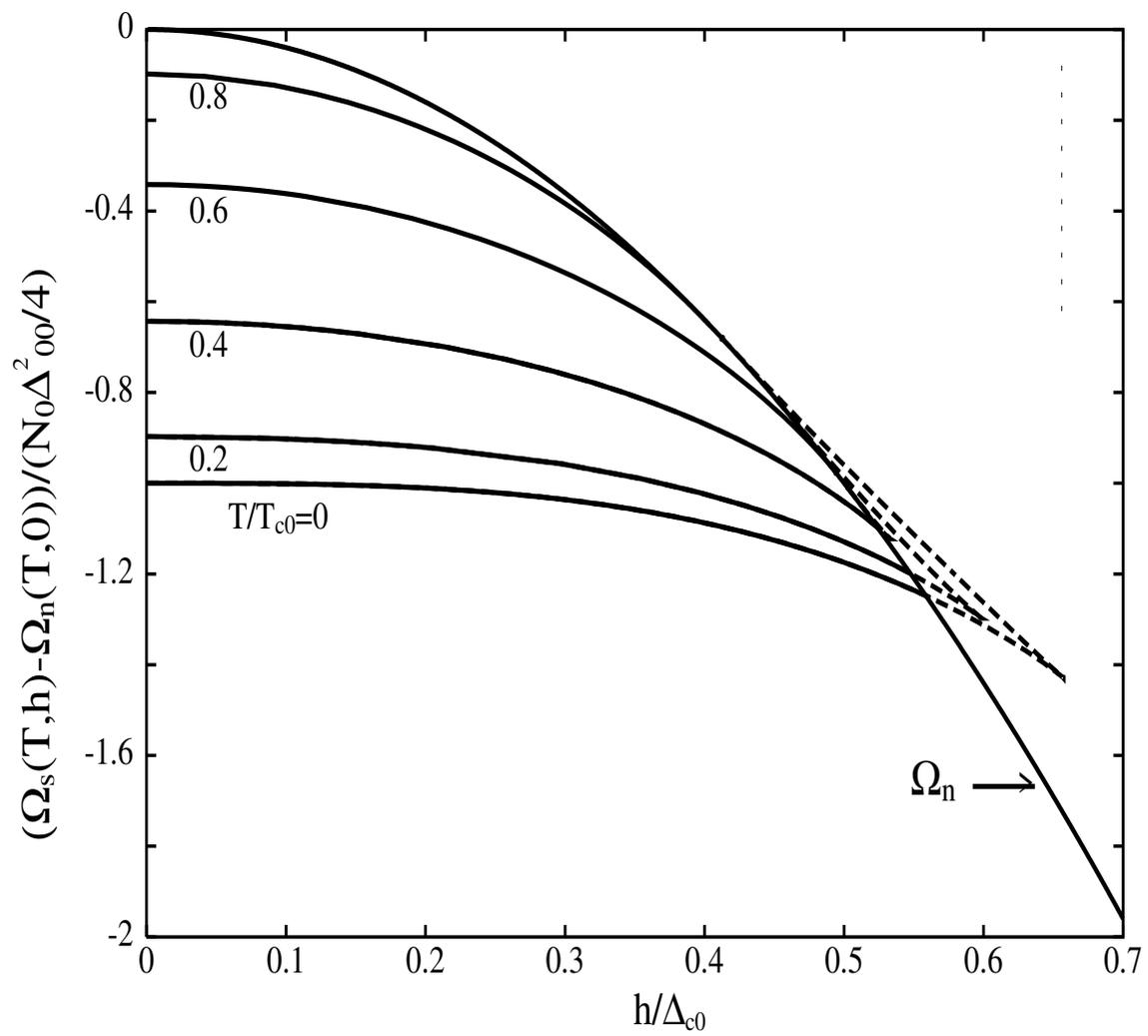
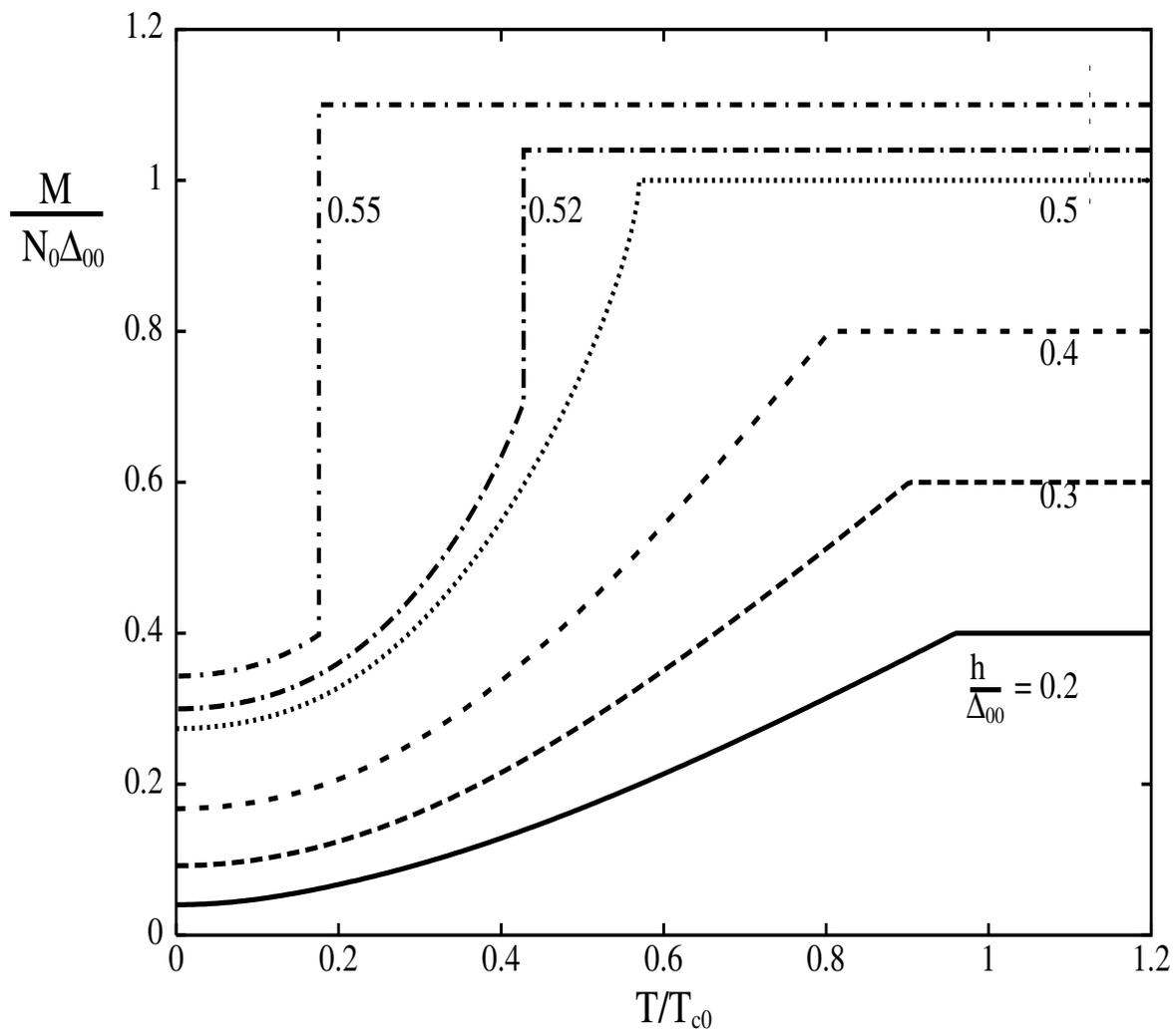
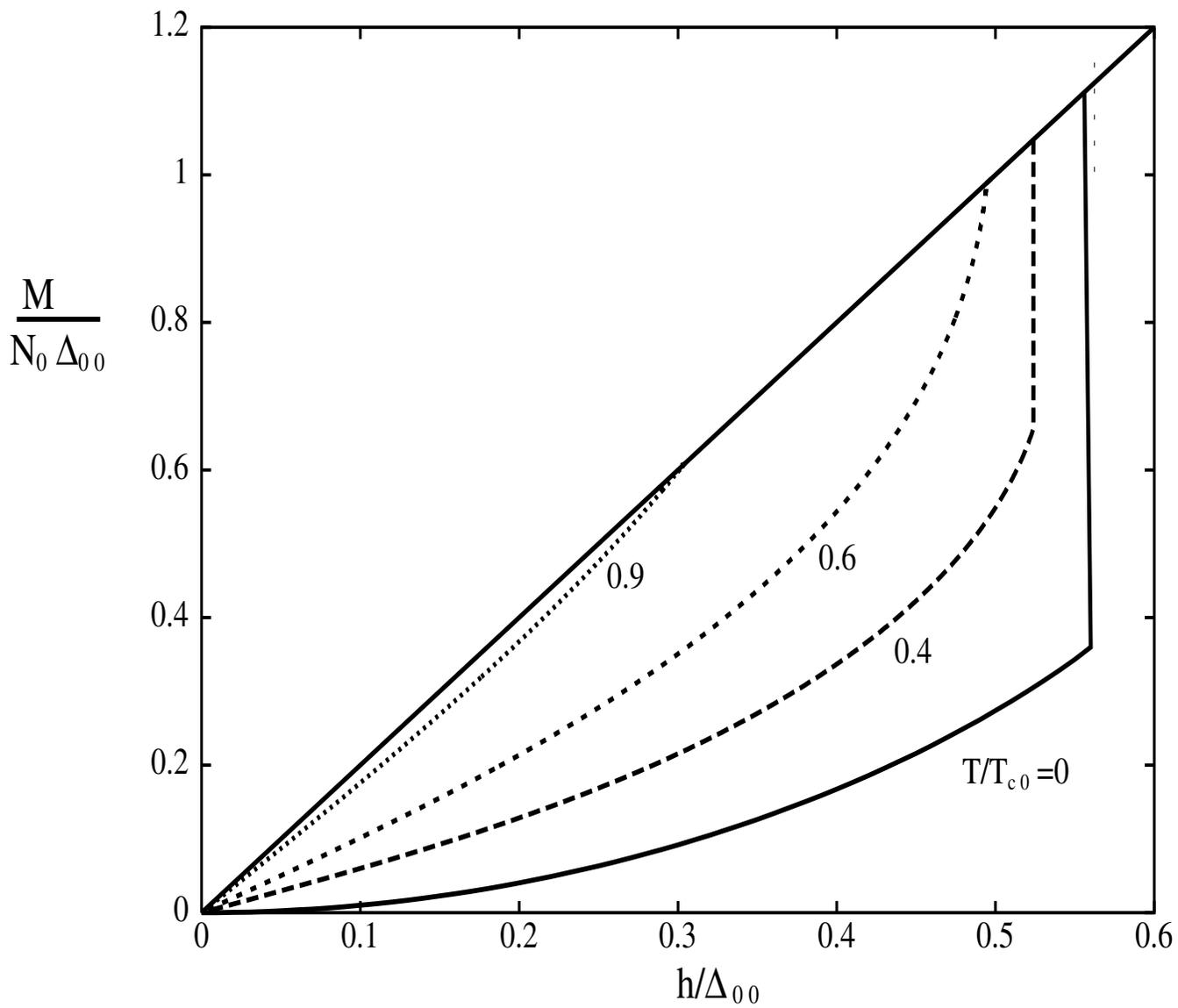
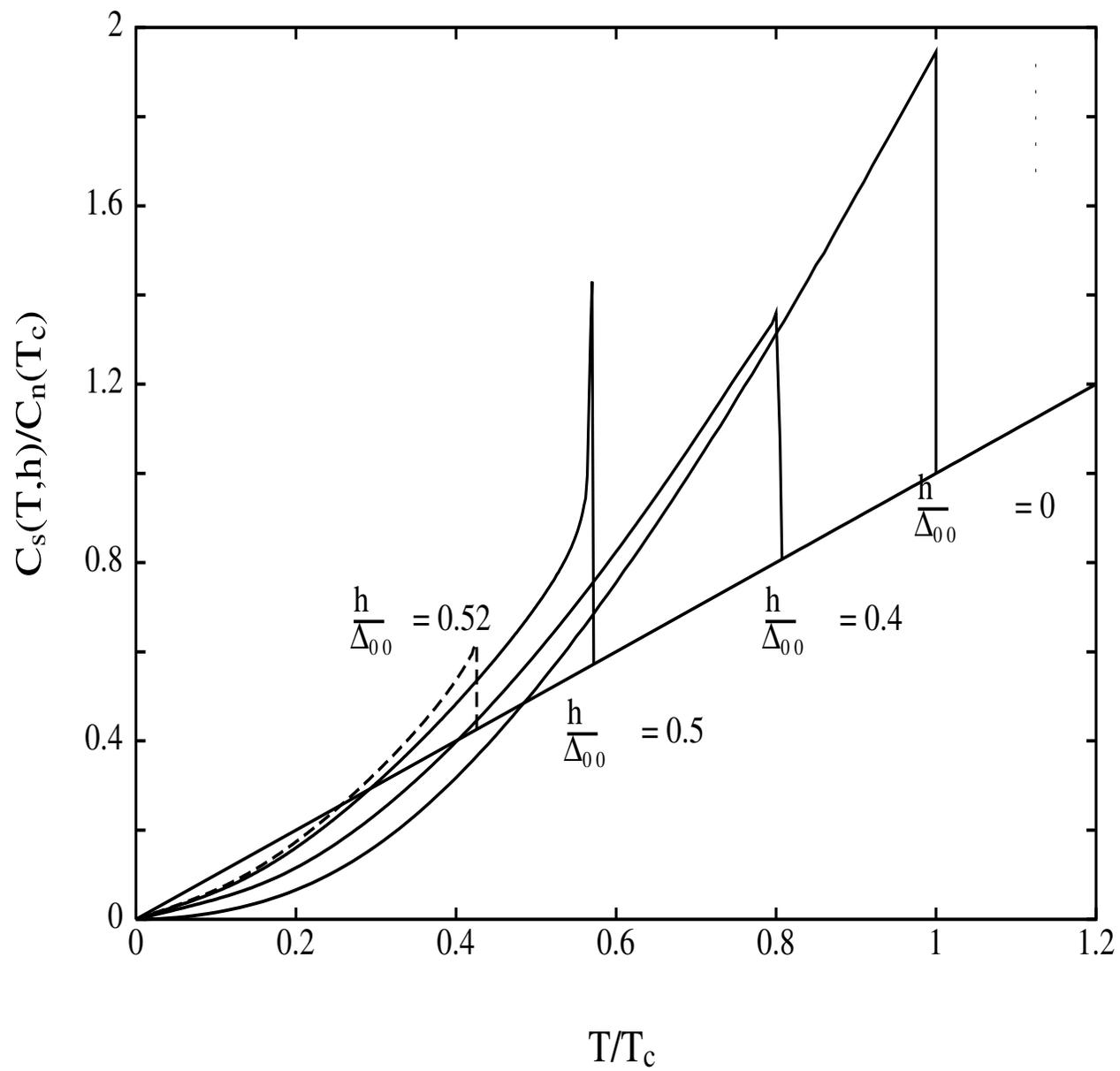
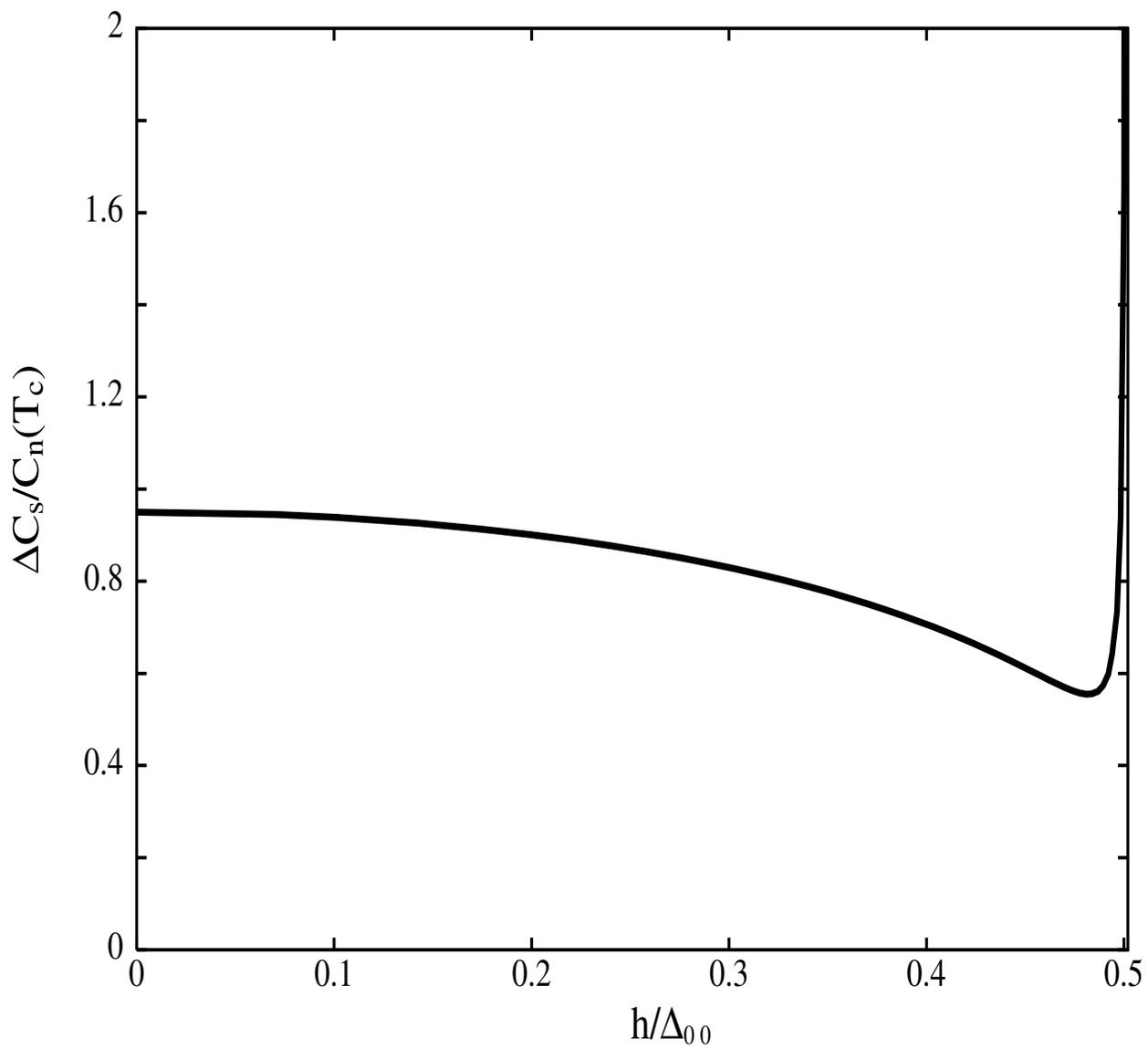
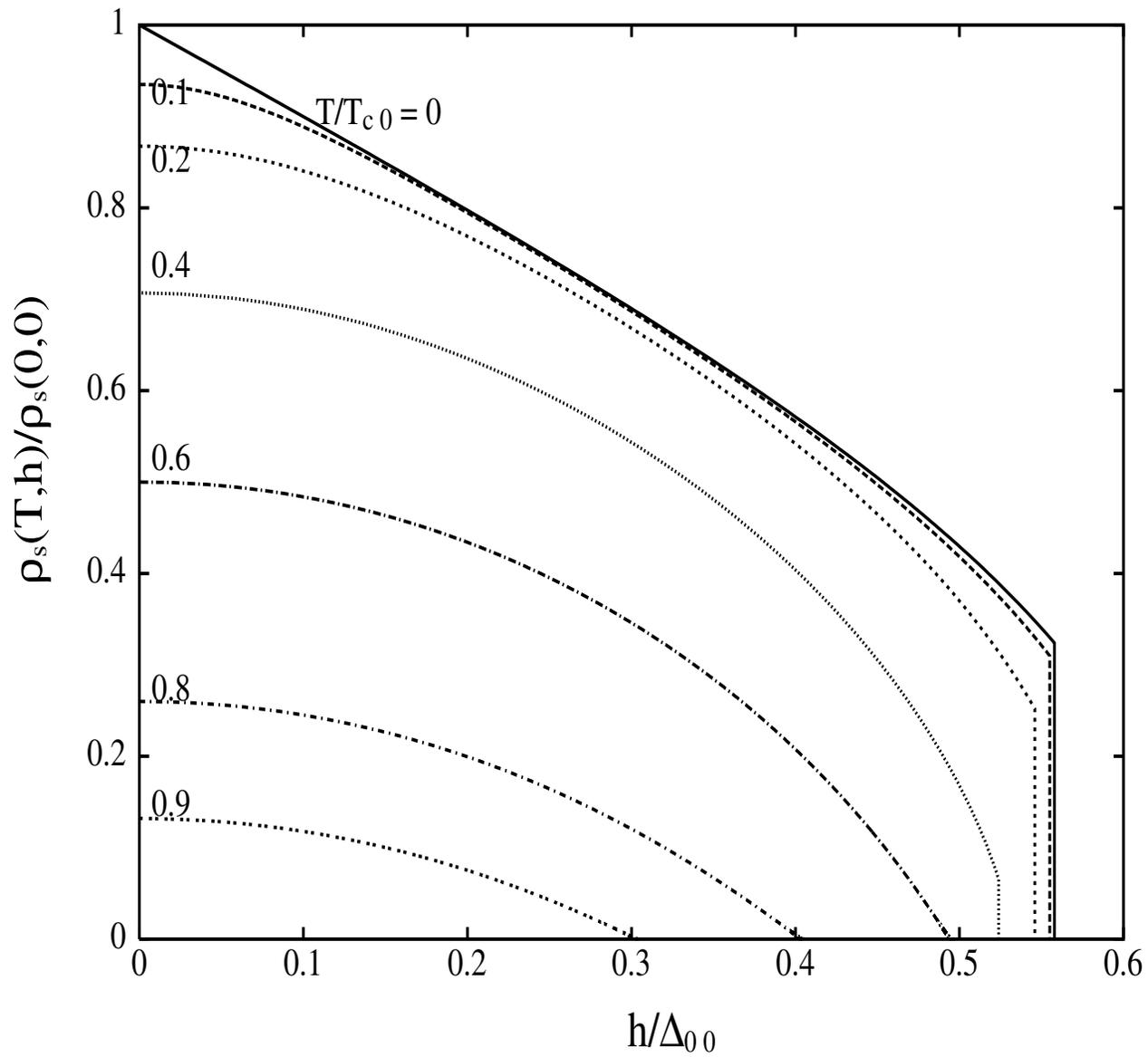
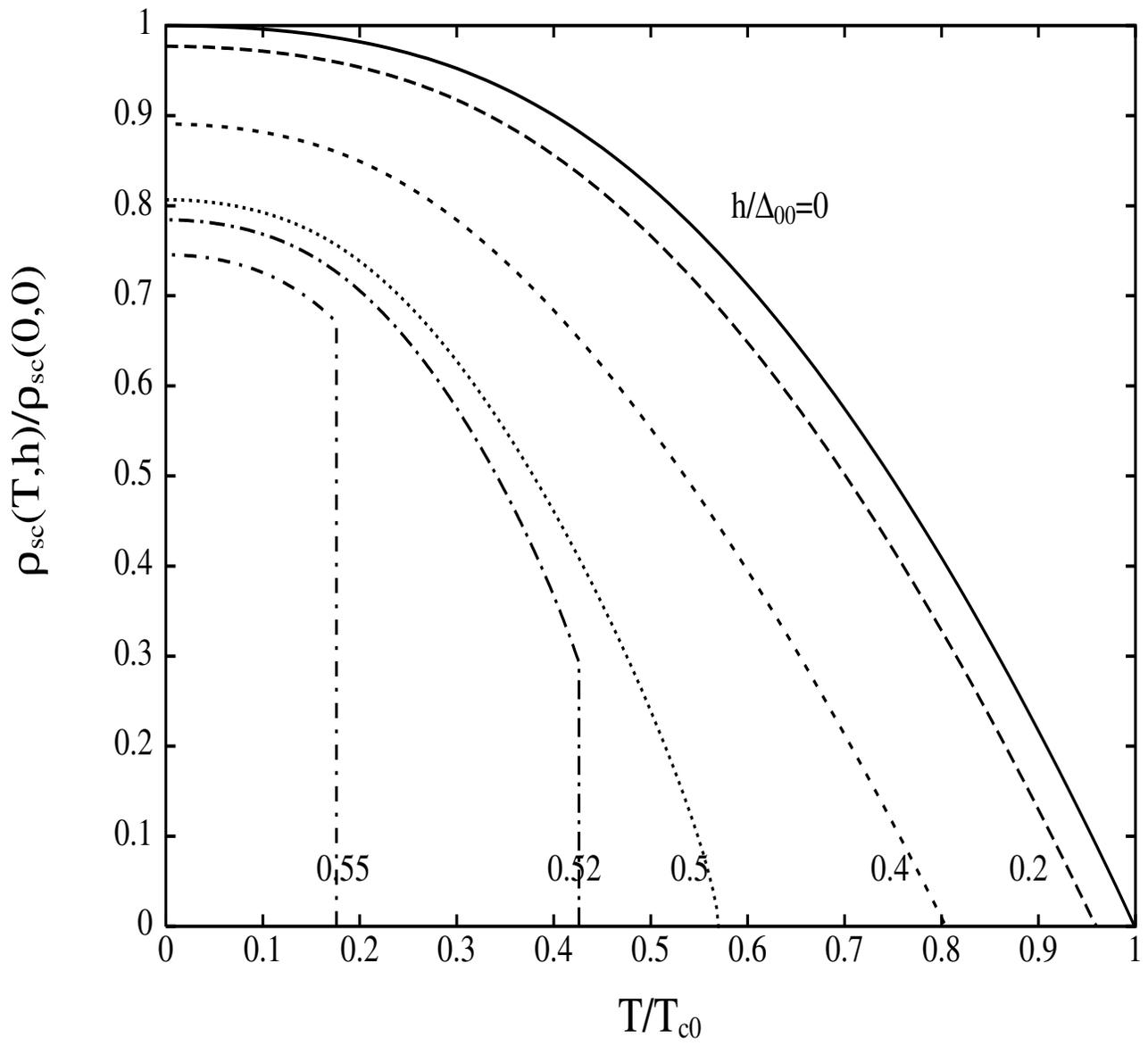
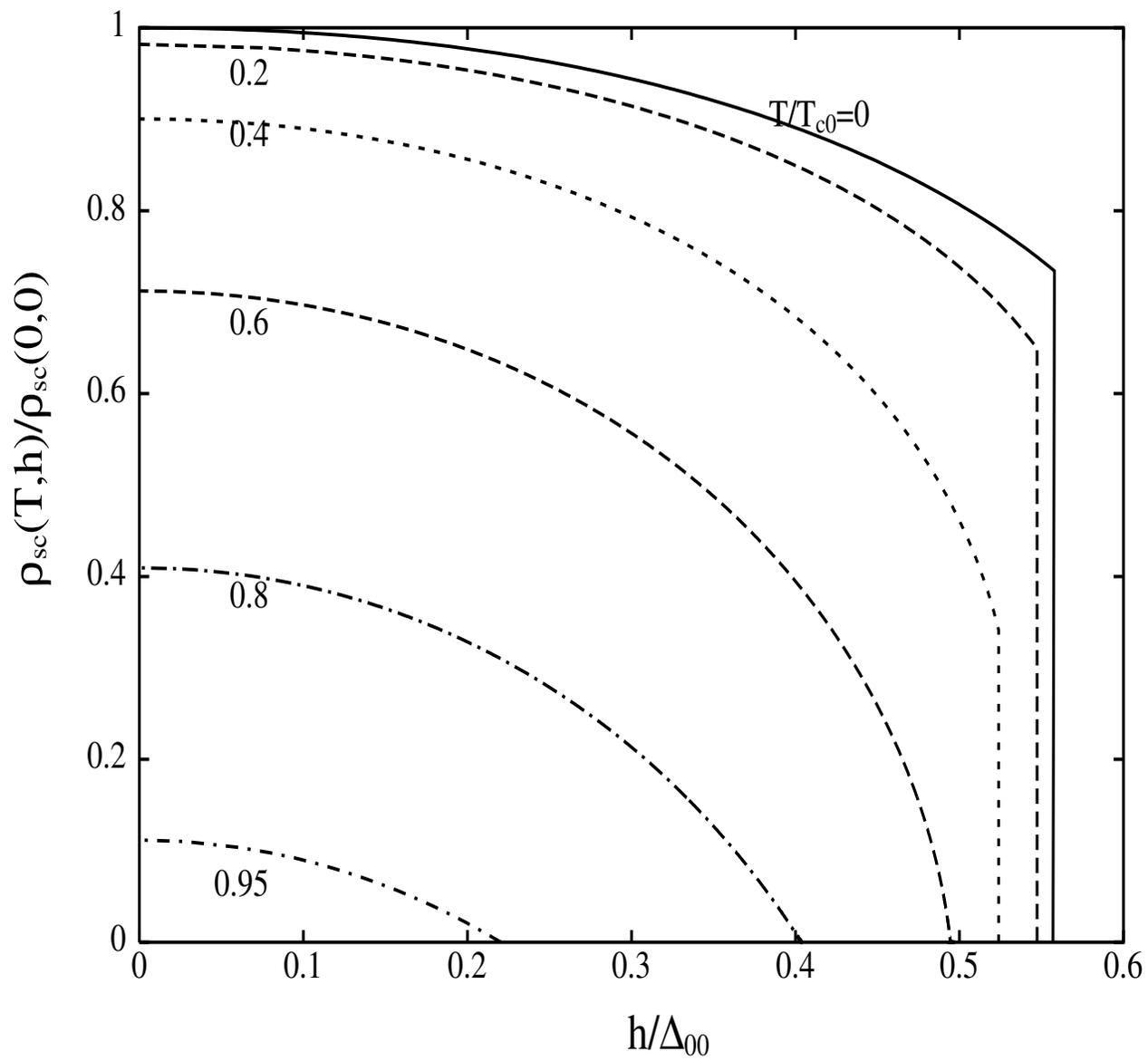
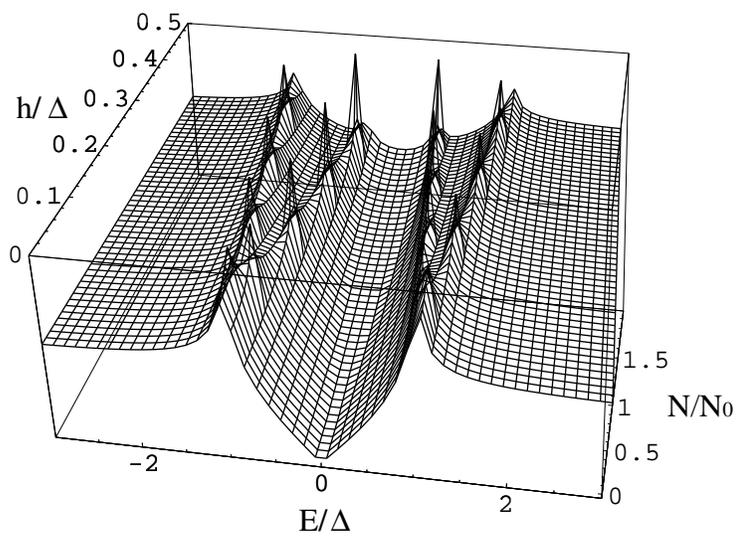
\end{document}